\RequirePackage[mathlines]{lineno} 
\documentclass[a4paper,11pt]{article}

\usepackage{jheppub} 
\usepackage{overpic}
\usepackage{amssymb}
\usepackage{amsmath}
\usepackage[T1]{fontenc} 
\usepackage{besphysics}

\usepackage{amsmath}
\usepackage{graphicx}
\usepackage{subfigure}
\usepackage{epstopdf}

\usepackage{threeparttable}
\usepackage{multirow}
\usepackage{subfigure}
\usepackage{float}
\let\oldequation\equation
\let\oldendequation\endequation
\renewenvironment{equation}
  {\linenomathNonumbers\oldequation}
  {\oldendequation\endlinenomath}

\begin{document}

\title{\bf \boldmath
Observation of the Doubly Cabibbo-Suppressed Decays $D^+\to K^+\pi^0\pi^0$ and $D^+\to K^+\pi^0\eta$
}
\collaborationImg{\includegraphics[height=4cm,angle=90 ]{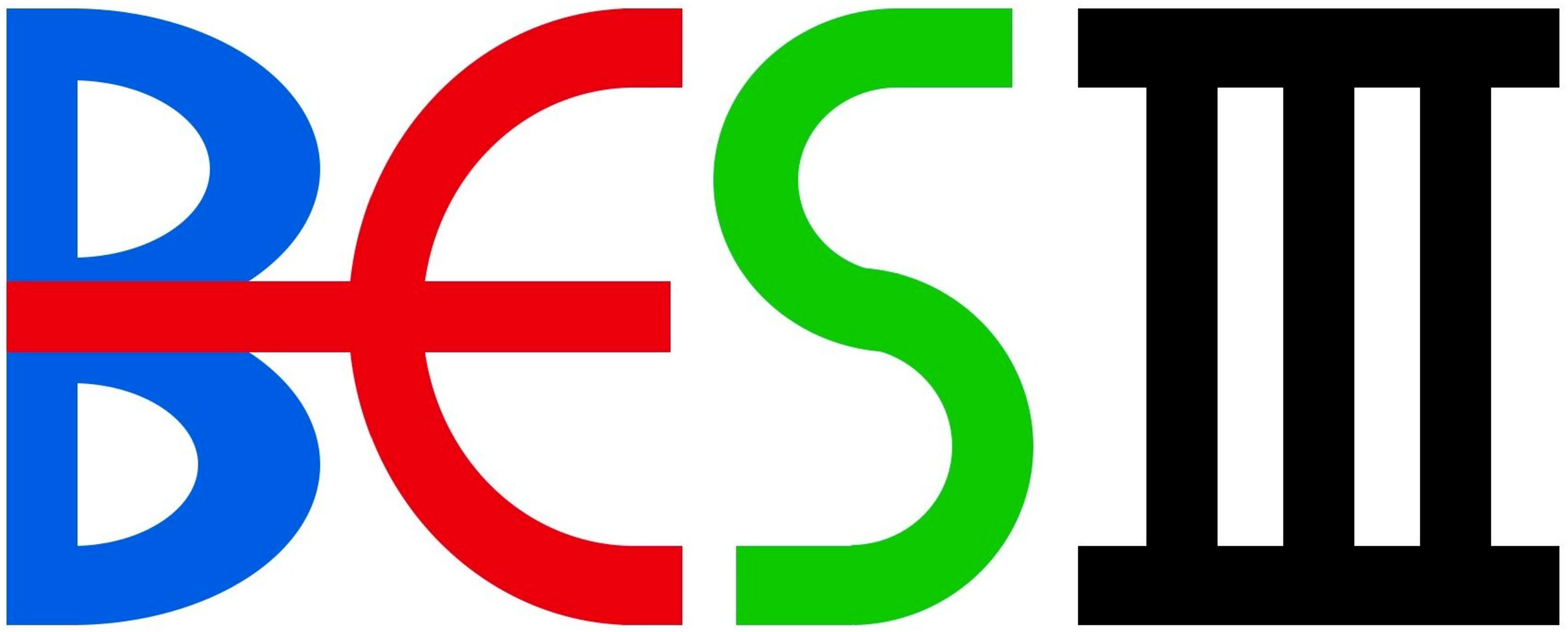}}

\collaboration{BESIII Collaboration}
\author{
M.~Ablikim$^{1}$, M.~N.~Achasov$^{10,b}$, P.~Adlarson$^{68}$, S. ~Ahmed$^{14}$, M.~Albrecht$^{4}$, R.~Aliberti$^{28}$, A.~Amoroso$^{67A,67C}$, M.~R.~An$^{32}$, Q.~An$^{64,50}$, X.~H.~Bai$^{58}$
% ...
\\
\vspace{0.2cm}
(BESIII Collaboration)\\
\vspace{0.2cm} {\it
$^{1}$ Institute of High Energy Physics, Beijing 100049, People's Republic of China\\
% ...
$^{4}$ Bochum Ruhr-University, D-44780 Bochum, Germany\\
% ...
$^{10}$ G.I. Budker Institute of Nuclear Physics SB RAS (BINP), Novosibirsk 630090, Russia\\
% ...
$^{14}$ Helmholtz Institute Mainz, Staudinger Weg 18, D-55099 Mainz, Germany\\
% ...
$^{28}$ Johannes Gutenberg University of Mainz, Johann-Joachim-Becher-Weg 45, D-55099 Mainz, Germany\\
% ...
$^{32}$ Liaoning Normal University, Dalian 116029, People's Republic of China\\
% ...
$^{50}$ State Key Laboratory of Particle Detection and Electronics, Beijing 100049, Hefei 230026, People's Republic of China\\
% ...
$^{58}$ University of Jinan, Jinan 250022, People's Republic of China\\
% ...
$^{64}$ University of Science and Technology of China, Hefei 230026, People's Republic of China\\
% ...
$^{67}$ University of Turin and INFN, (A)University of Turin, I-10125, Turin, Italy; (B)University of Eastern Piedmont, I-15121, Alessandria, Italy; (C)INFN, I-10125, Turin, Italy\\
$^{68}$ Uppsala University, Box 516, SE-75120 Uppsala, Sweden\\
% ...
$^{b}$ Also at the Novosibirsk State University, Novosibirsk, 630090, Russia\\
% ...
}
}

\abstract{
By analyzing $e^+e^-$ annihilation data corresponding to an integrated luminosity of $2.93\,\rm fb^{-1}$ collected
at the center-of-mass energy of 3.773\,GeV with the BESIII detector, we report the first observations of the doubly Cabibbo-suppressed decays $D^+\to K^+\pi^0\pi^0$ and $D^+\to K^+\pi^0\eta$. The branching fractions of $D^+\to K^+\pi^0\pi^0$ and $D^+\to K^+\pi^0\eta$ are measured to be $(2.1 \pm 0.4_{\rm stat} \pm 0.1_{\rm syst})\times 10^{-4}$ and $(2.1 \pm 0.5_{\rm stat} \pm 0.1_{\rm syst})\times 10^{-4}$ with statistical significances of 8.8$\sigma$ and 5.5$\sigma$, respectively.
In addition, we search for the subprocesses $D^+\to K^{*}(892)^{+}\pi^0$ and $D^+\to K^{*}(892)^{+}\eta$ with $K^{*}(892)^+\to K^+\pi^0$.
The branching fraction of $D^+\to K^{*}(892)^{+}\eta$ is determined to be $({4.4^{+1.8}_{-1.5}}_{\rm stat}\pm0.2_{\rm syst})\times10^{-4}$,
with a statistical significance of 3.2$\sigma$. No significant signal for $D^+\to K^{*}(892)^{+}\pi^0$ is found
and we set an upper limit on the branching fraction of this decay at the 90\% confidence level to be $5.4\times10^{-4}$.
}

\maketitle
\flushbottom

\section{INTRODUCTION}

Hadronic $D$ decays open an important window to explore weak $D$ decay mechanisms.
Based on SU(3)-flavor symmetry, the branching fractions (BFs) of two-body hadronic $D\to VP$ decays, where $V$ and $P$ denote vector and pseudoscalar mesons, have been calculated with various approaches~\cite{yufs,theory_2,theory_3}.
The effect of SU(3)-flavor symmetry breaking has been validated in Cabbibo-favored (CF) and singly Cabibbo-suppressed $D\to VP$ decays.
However, experimental information related to doubly Cabibbo-suppressed (DCS) $D\to VP$ decays is rare, due to their small BFs coupled with large backgrounds.
The BFs of the DCS decays $D^+\to K^{*+}\pi^0$ and $D^+\to K^{*+}\eta$ are predicted to be $\sim10^{-4}$,
and the ratio of these branching ratios $\frac{{\mathcal B}_{D^+\to K^{*+}\pi^0}}{{\mathcal B}_{D^+\to K^{*+}\eta}}$ is estimated
 to be either $2.86\pm0.76$~\cite{yufs} or $4$~\cite{theory_2}.
Improved understanding of U-spin and SU(3)-flavor
symmetry breaking effects can be derived from these decays, which can lead to more precise theoretical predictions of $CP$ violation in the charm sector~\cite{yufs,ref5,theory_a,theory_5,theory_4,theory_3,theory_2,theory_1}.

Unlike the DCS decay $D^+\to K^0\pi^+\pi^0$, which has a large irreducible background from the CF decay $D^+\to \bar K^0\pi^+\pi^0$, the $D^+\to K^+\pi^0\pi^0$ decay offers a unique low-background opportunity to investigate $D^+\to K^{*+}\pi^0$ with $K^{*+}\to K^+\pi^0$ decay. A similar argument can be made to study $D^+\to K^+\pi^0\eta$ decays.
Isospin statistical models indicate that the BF of $D^+\to K^+\pi^0\pi^0$ is one-third of that of $D^+\to K^+\pi^+\pi^-$~\cite{sim1,sim2}. Since the BF of the DCS decay $D^+\to K^+\pi^+\pi^-$ relative to its CF counterpart $D^+\to K^-\pi^+\pi^+$ is naively expected to be about $2\tan^4\theta_C$~\cite{pdg2020}, where $\tan^4\theta_C=0.29\%$ and $\theta_C$ is the Cabibbo mixing angle~\cite{cabibbo-angle,tan8},
the ratio $\frac{{\mathcal B}_{D^+\to K^+\pi^0\pi^0}} {{\mathcal B}_{D^+\to K^-\pi^+\pi^+}}$ is expected to be $\frac{2}{3}\tan^4\theta_C$.
Therefore, experimental studies of $D^+\to K^+\pi^0\pi^0$ and  $D^+\to K^+\pi^0\eta$ decays provide a powerful way to further understand the decay dynamics of charmed mesons.
Throughout the text, charge conjugated decays are always implied and $K^{*+}$ denotes the $K^{*}(892)^+$, which has a mass of 0.892 GeV/$c^2$~\cite{pdg2020}.

This paper reports the first experimental studies of $D^+\to K^+\pi^0\pi^0$, $D^+\to K^+\pi^0\eta$, $D^+\to K^{*+}\pi^0$, and $D^+\to K^{*+}\eta$.
This analysis uses a sample of $e^+e^-$ annihilation data~\cite{lum_bes31,lum_bes32}
taken with the BESIII detector at the center-of-mass energy $\sqrt s=$ 3.773~GeV.
This energy point is above the threshold to produce $D\bar D$ and below that to produce $D^*\bar D$, where $D$ and $D^*$ denote charged or neutral charmed meson and their excited states, respectively. Therefore, the $D$ and $\bar D$ mesons are produced exclusively in pairs, with no additional hadrons accompanying them.  This sample corresponds to an integrated luminosity of 2.93\,fb$^{-1}$.

\section{BESIII DETECTOR AND MONTE CARLO SIMULATION}

The BESIII detector is a magnetic spectrometer~\cite{BESIII} located at the Beijing Electron
Positron Collider (BEPCII)~\cite{Yu:IPAC2016-TUYA01}. The
cylindrical core of the BESIII detector consists of a helium-based
 multilayer drift chamber (MDC), a plastic scintillator time-of-flight
system (TOF), and a CsI (Tl) electromagnetic calorimeter (EMC),
which are all enclosed in a superconducting solenoidal magnet
providing a 1.0~T magnetic field. The solenoid is supported by an
octagonal flux-return yoke with resistive-plate counter muon-identifier modules interleaved with steel. The acceptance of
charged particles and photons is 93\% over $4\pi$ solid angle. The
charged-particle momentum resolution at $1~{\rm GeV}/c$ is
$0.5\%$, and the resolution of the specific ionization energy loss ($dE/dx$) is $6\%$ for the electrons
from Bhabha scattering. The EMC measures photon energies with a
resolution of $2.5\%$ ($5\%$) at $1$~GeV in the barrel (end cap)
region. The time resolution of the TOF barrel part is 68~ps, while
that of the end cap part is 110~ps.

Details about the design and performance of the BESIII detector are given in Refs.~\cite{BESIII}.
Simulated samples produced with a {\sc
Geant4}-based~\cite{geant4,geant4-1,geant4-2} Monte Carlo (MC) simulation, which
includes the geometric description of the BESIII detector and the
detector response, are used to determine the detection efficiency
and to estimate backgrounds. The simulation includes the beam
energy spread and initial state radiation (ISR) in the $e^+e^-$
annihilations modeled with the generator {\sc
kkmc}~\cite{kkmc1,kkmc2}.
The signal of $D^+\to K^+\pi^0\pi^0(\eta)$ is simulated using an MC generator that incorporates
the resonant decay $D^+\to K^{*+}\pi^0(\eta)$ and the phase space decay $D^+\to K^+\pi^0\pi^0(\eta)$.
The background is studied using an inclusive MC sample that consists of the
production of $D\bar D$ pairs with consideration of quantum coherence for all neutral $D$
modes, the non-$D\bar D$ decays of the $\psi(3770)$, the ISR
production of the $J/\psi$ and $\psi(3686)$ states, and the
continuum processes incorporated in {\sc kkmc}~\cite{kkmc1,kkmc2}.
The known decay modes are modeled with {\sc
evtgen}~\cite{evtgen1,evtgen2} using the corresponding BFs taken from the
Particle Data Group~\cite{pdg2020}, while the remaining unknown decays
from the charmonium states are modeled with {\sc
lundcharm}~\cite{lundcharm,lundcharm1}. Final state radiation
from charged final state particles is incorporated using {\sc photos}~\cite{photos,photos-1,photos-2}.

\section{MEASUREMENT METHOD AND SINGLE TAG YIELDS}

The BFs of the signal decays are measured with a double-tag technique that was first developed by the Mark III Collaboration~\cite{mark3}.
The signal $D^+$ decays are reconstructed alongside hadronic $D^-$ decays to $K^+\pi^-\pi^-$, $K^0_S\pi^-$ and $K^+\pi^-\pi^-\pi^0$. This tag combination is chosen from the
six widely-used $D^-$ tag modes of $D^-\to K^+\pi^-\pi^-$, $K^0_S\pi^-$, $K^+\pi^-\pi^-\pi^0$, $K^0_S\pi^-\pi^0$, $K^0_S\pi^+\pi^+\pi^-$ and $K^+K^-\pi^-$ in most studies of $D^+$ decays, 
based on the optimization of the figure of merit $S/\sqrt {S+B}$.
Here, $S$ is the signal yield expected based on the known BFs of $D^+\to K^+\pi^+\pi^-$ or $D^+\to K^0_S\pi^+\eta$,
which are isospin symmetric decays of the DCS decays of interest;
and $B$ is the scaled background yield estimated by the inclusive MC sample.
The fully reconstructed $D^-$ is called the single-tag~(ST) meson.
Events in which both the signal $D^+$ meson and the ST $D^-$ meson are found are called double-tag~(DT) events.
For a given signal decay, the decay BF is determined by
\begin{equation}
\label{eq:br}
{\mathcal B}_{{\rm sig}} = N_{\rm DT}/({N_{\rm ST}\cdot \epsilon_{\rm sig}\cdot {\mathcal B}_{{\rm sub}}}),
\end{equation}
where ${N_{\rm ST}}$ and $N_{\rm DT}$ are the yields of ST and DT candidates in data,
$\epsilon_{\rm sig}=\sum\limits_{i=1}^{3}[(N_{\rm ST}^i\cdot \epsilon_{\rm DT}^i)/(N_{\rm ST}\cdot \epsilon_{\rm ST}^i)]$ is the signal efficiency in the presence of the ST candidate,
in which $\epsilon_{\rm ST}$ and $\epsilon_{\rm DT}$ are the efficiencies of selecting ST and DT candidates,
and $i$ stands for tag modes.
The ${\mathcal B}_{{\rm sub}}$ is the product of branching fractions of the subdecays of $K^{*+}$, $\pi^0$ and $\eta$.

Candidate $K^0_S$, $\pi^0$, and $\eta$ mesons are formed via the decays $K^0_S\to\pi^+\pi^-$, $\pi^0\to\gamma\gamma$, and $\eta\to\gamma\gamma$.
The $K^\pm$, $\pi^\pm$, $K^0_S$, $\pi^0$, and $\eta$ candidates are reconstructed and identified using the same
criteria as in
Refs.~\cite{bes3-etaetapi,papnew2}.

The ST $D^-$ mesons are distinguished from combinatorial background using two kinematic variables: the energy difference $\Delta E_{\rm tag} \equiv E_{D^-} - E_{\rm b}$ and the beam-constrained mass $M_{\rm BC}^{\rm tag} \equiv \sqrt{E^{2}_{\rm b}-|\vec{p}_{D^-}|^{2}}$.
Here, $E_{\rm b}$ is the beam energy,
and $\vec{p}_{D^-}$ and $E_{D^-}$ are the momentum and energy, respectively, of the $D^-$ candidate in the rest frame of the $e^+e^-$ system.
If more than one candidate survives the selection criteria of a given tag mode, the combination with the minimum $|\Delta E_{\rm tag}|$ is chosen.
Tagged $D^-$ candidates are selected with a requirement of $\Delta E_{\rm tag}\in(-25,\, 25)$\,MeV to suppress combinatorial backgrounds
in the $M_{\rm BC}^{\rm tag}$ distributions.
To extract the number of ST $D^-$ mesons for each tag mode, maximum likelihood fits have been performed on the individual $M_{\rm BC}^{\rm tag}$
distributions~\cite{bes3-etaetapi,papnew2}.
The ST yields and efficiencies for various tag modes are summarized in Table~\ref{tab:doubletageff2pipm}.
The number of ST $D^-$ mesons summed over the three tag modes is ${N_{{\rm ST}}=}\ (1150.3\pm1.5_{\rm stat})\times 10^3$.

\section{YIELDS OF DOUBLE-TAG EVENTS}

Candidates for the DCS $D^+$ decays are selected with the residual neutral and charged particles not used in the $D^-$ tag reconstruction.
Similar to the tag side, the energy difference and beam-constrained mass of the signal side,
$\Delta E_{\rm sig}$ and $M_{\rm BC}^{\rm sig}$, respectively, are calculated.
For each signal decay, if there are multiple combinations, the one giving the minimum $|\Delta E_{\rm sig}|$ is kept.
The accepted candidates are required to fall in the intervals $\Delta E_{\rm sig}\in (-78,36)\,{\rm MeV}$ and $\Delta E_{\rm sig}\in (-52,31)\,{\rm MeV}$ for $D^+\to K^+\pi^0\pi^0$ and $D^+\to K^+\pi^0\eta$, respectively.
To reduce background events from non-$D^+D^-$ processes,
the minimum opening angle between the $D^+$ and $D^-$ must be greater than $167^\circ$.
This requirement suppresses 57\%\,(81\%) of background for $D^+\to K^+\pi^0\pi^0(\eta)$ at the cost of losing 9\% of the two signal decays.
For $D^+\to K^+\pi^0\pi^0$,  the invariant mass of the $\pi^0\pi^0$ combination is required to be outside $(0.388,0.588)$ GeV/$c^2$ to reject the dominant background from the singly Cabibbo-suppressed decay $D^+\to K^+K_S^0(\to \pi^0\pi^0)$.

\begin{figure}[tp]
  \centering
\includegraphics[width=1\linewidth]{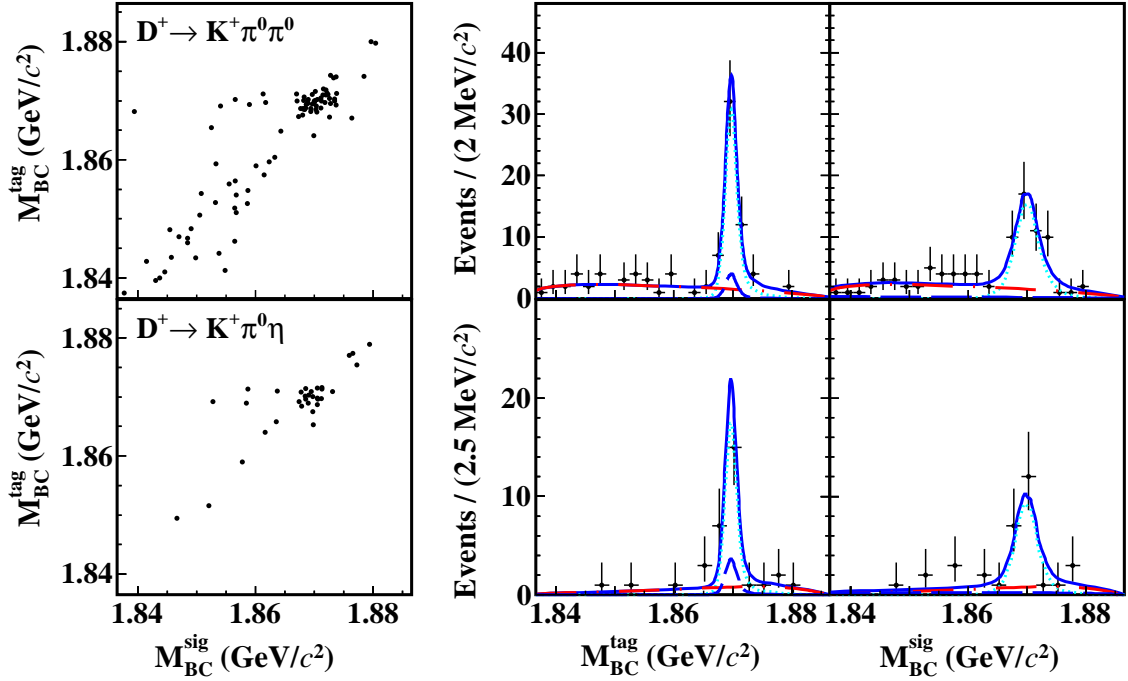}
  \caption{\small
Distributions of (left column) $M_{\rm BC}^{\rm tag}$ versus $M_{\rm BC}^{\rm sig}$  and
the projections on (middle column) $M_{\rm BC}^{\rm tag}$ and (right column) $M_{\rm BC}^{\rm sig}$ of the 2D fits to the DT candidate events.
The top row is for $D^+\to K^{+}\pi^0\pi^0$ and the bottom row is for $D^+\to K^{+}\pi^0\eta$.
Points with error bars are data.
Blue solid curves are the fit results.
Cyan dotted curves are the fitted signal distributions.
Blue long-dashed curves are BKGI.
Red dot-long-dashed curves are BKGII.}
\label{fig:2Dfit}
\end{figure}

The resulting distributions of $M_{\rm BC}^{\rm tag}$ versus $M_{\rm BC}^{\rm sig}$ of the accepted DT candidates are shown in the left column of Fig.~\ref{fig:2Dfit}.
Signal events cluster around $M_{\rm BC}^{\rm tag} = M_{\rm BC}^{\rm sig} = M_{D^+}$,
where $M_{D^+}$ is the known $D^+$ mass~\cite{pdg2020}.
There are three kinds of background events.
The events with correctly reconstructed $D^+$ ($D^-$) and incorrectly reconstructed $D^-$ ($D^+$) are called BKGI.
These background events are distributed along the horizontal and vertical bands around the known $D^+$ mass.
The events spreading along the diagonal, which are mainly from the $e^+e^- \to q\bar q$ processes, are named BKGII.
The events with incorrectly reconstructed $D^-$ and $D^+$
are dispersed in the allowed kinematic region and they are ignored in the following analysis due to  limited statistics.

The signal yields of the DT events are extracted from a two-dimensional (2D) unbinned maximum likelihood fit
to the corresponding distribution of $M_{\rm BC}^{\rm tag}$ versus $M_{\rm BC}^{\rm sig}$.
The signal shape is described by the 2D probability density function (PDF) from the MC simulation after convolving with a Gaussian resolution function with parameters derived from the control sample of $D^+\to \pi^+\pi^0\pi^0$.
For various background components, the individual PDFs are constructed as~\cite{cleo-2D,papnew2}
\begin{itemize}
\item BKGI: $b(x)\cdot c_y(y;E_{\rm b},\xi_{y})$,
\item BKGII: $c_z(z;\sqrt{2}E_{\rm b},\xi_{z}) \cdot g(k;0,\sigma_k)$,
\end{itemize}
Here, $x=M_{\rm BC}^{\rm tag}$, $y=M_{\rm BC}^{\rm sig}$, $z=(x+y)/\sqrt{2}$, and $k=(x-y)/\sqrt{2}$.
The one-dimensional MC-simulated signal shape is $b(x)$.
The $c_f$ is an ARGUS function~\cite{ARGUS} defined as
\begin{equation}
c_f\left(f; E_{\rm end}, \xi_f\right) = A_f \cdot f (1 - \frac {f^2}{E_{\rm end}^2})^{\frac{1}{2}} \cdot e^{\xi_f \cdot (1-\frac {f^2}{E_{\rm end}^2})},
\end{equation}
where $f \equiv$ $y$, or $z$,
$A_f$ is a normalization factor,
$\xi_f$ is a fit parameter, and
$E_{\rm end}$ is the endpoint fixed at $E_{\rm b}$ for $c_y$ or $\sqrt 2 E_{\rm b}$ for $c_z$.
The function $g(k;0,\sigma_k)$ is a Gaussian function with zero mean and standard deviation  $\sigma_k=\sigma_0 \cdot(\sqrt{2}E_{\rm b}-z)^p$,
where $\sigma_0$ and $p$ are the parameters determined from the fit.
All other parameters are free in the fit.
The spectra of the middle and right columns in Fig.~\ref{fig:2Dfit} show
the projections on $M_{\rm BC}^{\rm tag}$ and $M_{\rm BC}^{\rm sig}$ of the 2D fits to data.
These fits give the signal yields of $D^+\to K^+\pi^0\pi^0$ and $D^+\to K^+\pi^0\eta$ to be $42.8\pm7.2_{\rm stat}$
and $19.2\pm5.0_{\rm stat}$, respectively.

To account for the large difference of detection efficiencies between resonant and non-resonant decays,
we estimate the resonant component of $D^+\to K^{*+}\pi^0(\eta)$
under the assumption that the non-resonant component is uniformly distributed and there is no interference between the two kinds of components.
The signal yield of the resonant decay $D^+\to K^{*+}\pi^0(\eta)$ is extracted from
a simultaneous 2D fit in the $K^{*+}$ signal and sideband regions.

\begin{figure}[htp]
  \centering
\includegraphics[width=1.0\linewidth]{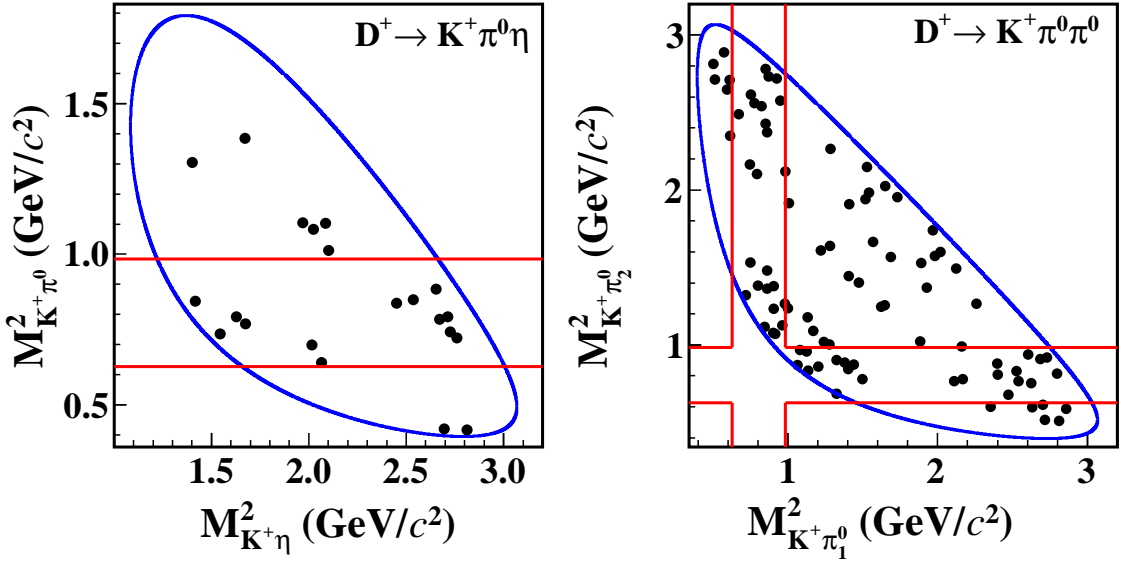}
  \caption{\small
  Distributions of (left) $M_{K^{+}\eta}^{2}$ versus $M_{K^{+}\pi^{0}}^{2}$ for $D^+\to K^+\pi^0\eta$ candidates and (right) $M_{K^{+}\pi^{0}_{1}}^{2}$ versus $M_{K^{+}\pi^{0}_{2}}^{2}$ for $D^+\to K^+\pi^0\pi^0$ candidates (two entries per event for symmetrization).
   In the kinematic region marked in blue, the regions inside and outside the red band are the $K^{*+}$ signal and sideband regions, respectively.
   The requirement of $|M_{\rm BC}^{\rm tag(sig)}- M_{D^+}|<0.005$ GeV/$c^2$ has been imposed.
  }
\label{Kband}
\end{figure}

The $K^{*+}$ signal region is defined as the invariant mass $M_{K^+\pi^0}\in(0.792,0.992)$ GeV/$c^2$ for $D^+\to K^{*+}\eta$ and one of two $M_{K^+\pi^0}$ combinations lying in $M_{K^+\pi^0}\in(0.792,0.992)$ GeV/$c^2$ for $D^+\to K^{*+}\pi^0$.
The sideband region is defined as the $K^+\pi^0$ combination
outside the $K^{*+}$ signal region but within the kinematic region.
Definitions of the $K^{*+}$ signal and sideband regions are shown in Fig.~\ref{Kband}.

\begin{figure*}[htbp]
\begin{center}
\includegraphics[width=0.495\linewidth]{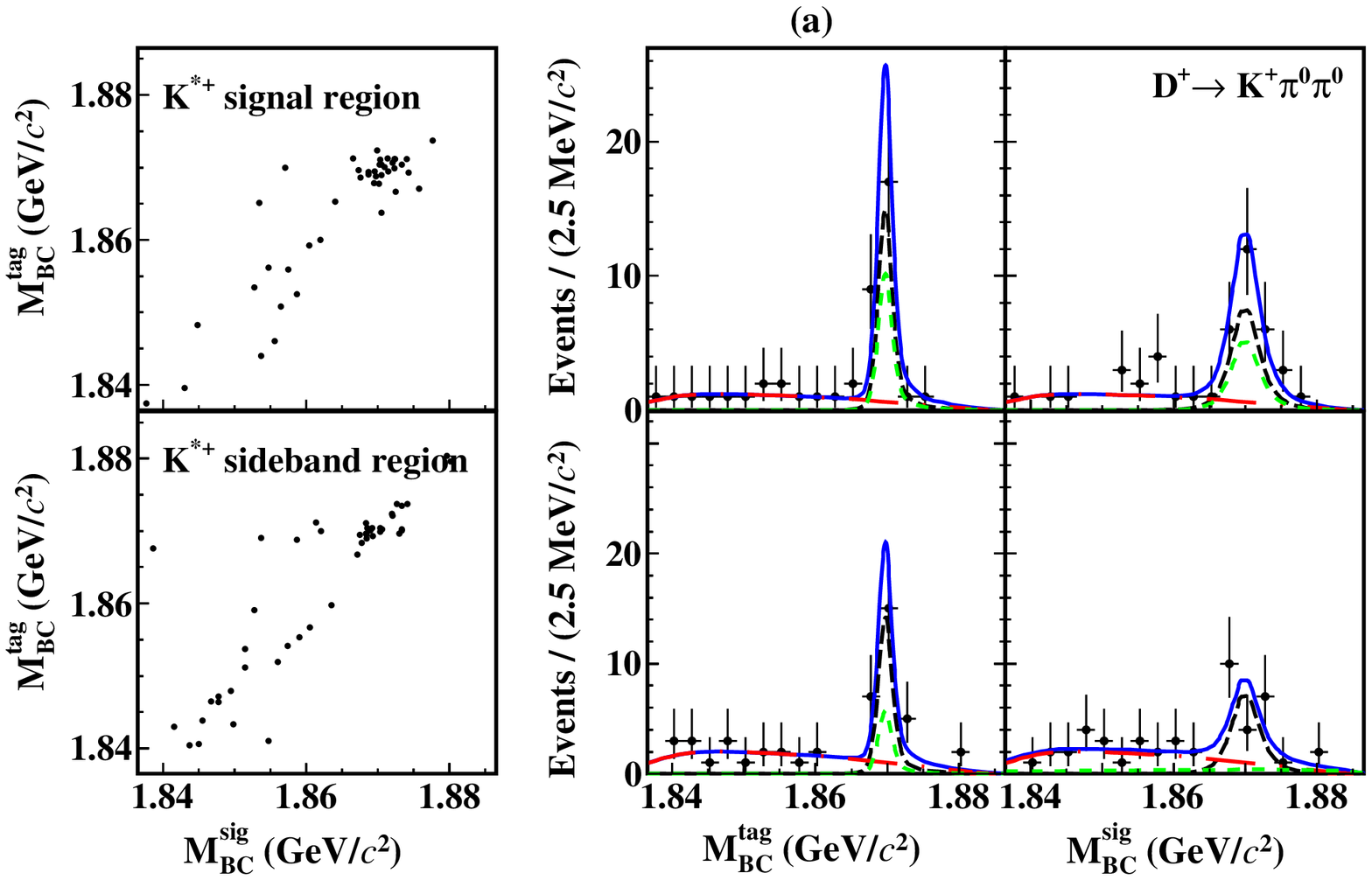}
\includegraphics[width=0.495\linewidth]{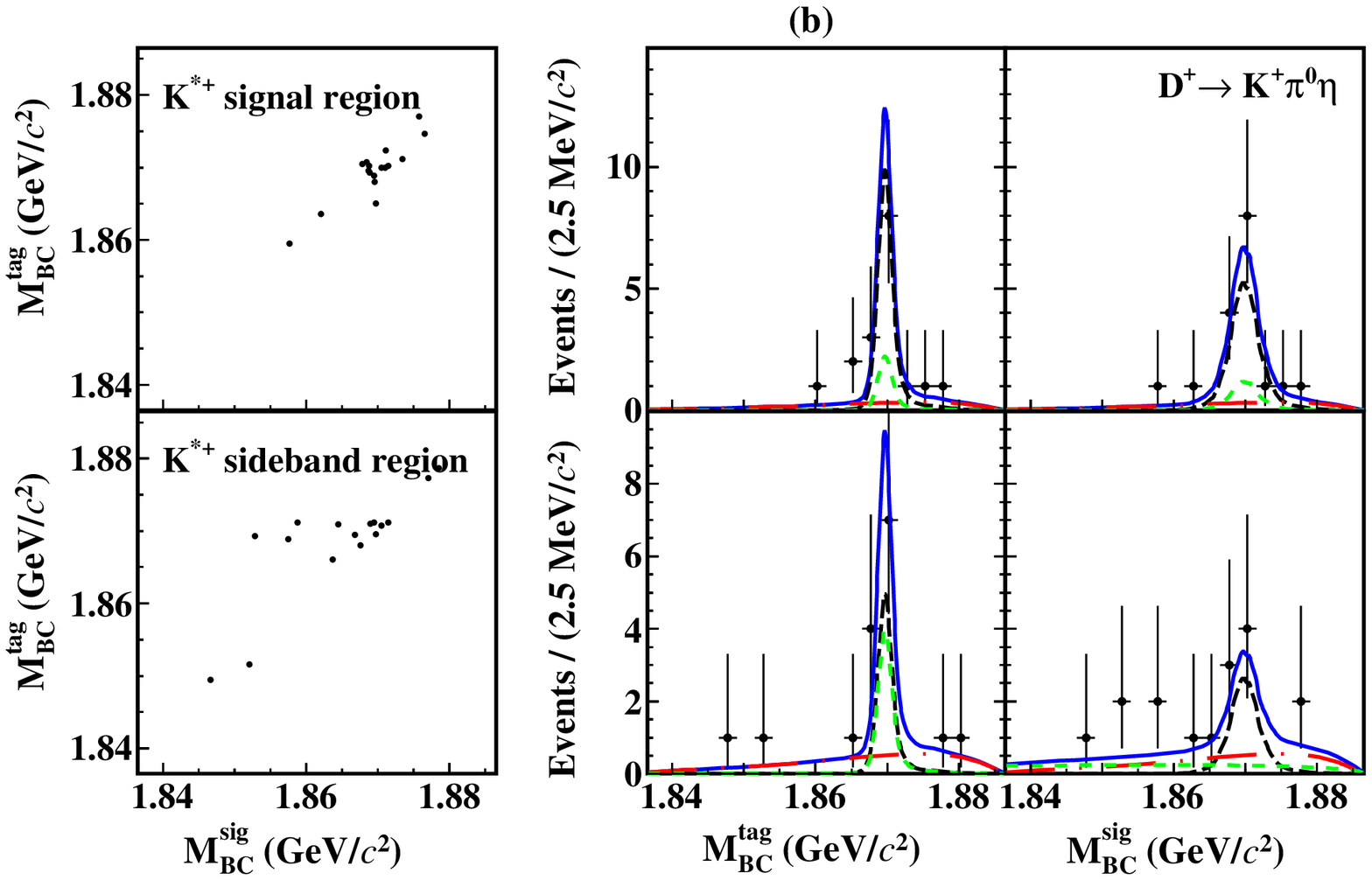}
\end{center}
\caption{
Distributions of (left column) $M_{\rm BC}^{\rm tag}$ versus $M_{\rm BC}^{\rm sig}$ and the projections on (middle column) $M_{\rm BC}^{\rm tag}$ and (right column) $M_{\rm BC}^{\rm sig}$ of the constrained 2D fits to the DT candidate events in the $K^{*+}$ signal region (top row) and sideband region (bottom row) for (a) $D^+\to K^+\pi^0\pi^0$ and (b) $D^+\to K^+\pi^0\eta$. Points with error bars are data.
Blue solid curves are the fit results.
Black dotted curves are the signal distributions.
For the $K^{*+}$ sideband region, green dotted and red dot-long-dashed curves are BKGI and BKGII, respectively.
For the $K^{*+}$ signal region, red dot-long-dashed curves are BKGII and green dotted curves are the peaking backgrounds constrained by using the $K^{*+}$ sideband events.
}
\label{fig:Dfit}
\end{figure*}

The left columns of Figs.~\ref{fig:Dfit}(a) and \ref{fig:Dfit}(b) show the $M_{\rm BC}^{\rm tag}$ versus $M_{\rm BC}^{\rm sig}$ distributions of
the accepted DT candidates, where the top and bottom rows correspond to the $K^{*+}$ signal and sideband regions, respectively.
In the simultaneous fits,
the ratios of the non-resonant background yields between the $K^{*+}$ sideband and signal regions
are fixed to the MC-determined values of $f_{K^{*+}\pi^0}=1.40\pm0.02$ for $D^+\to K^+\pi^0\pi^0$ and $f_{K^{*+}\eta}=2.25\pm0.05$ for $D^+\to K^+\pi^0\eta$, respectively, where the efficiency differences have been considered.
In addition, the parameters of the ARGUS functions in the 2D fit to the $K^{*+}$ sideband events are constrained to be the same as those for the $K^{*+}$ signal region.
The other parameters are left free.
These fits give the signal yields of $D^+\to K^{*+}\pi^0$ and $D^+\to K^{*+}\eta$ to be
${16.6^{+6.6}_{-6.2}}_{\rm stat}$ and ${10.9^{+4.4}_{-3.8}}_{\rm stat}$, respectively.
Combining the $D^+\to K^+\pi^0\pi^0$ and $D^+\to K^+\pi^0\eta$ signal yields,
we obtain the fractions of the resonant components to be $r_{K^{*+}\pi^0}= 0.39\pm0.17_{\rm stat}$ and $r_{K^{*+}\eta}= 0.57\pm0.28_{\rm stat}$, respectively.

The efficiency of detecting the signal decay $D^+\to K^+\pi^0\pi^0(\eta)$
is estimated by using a mixture of the signal MC events for the resonant decay $D^+\to K^{*+}\pi^0(\eta)$ and the phase space decay $D^+\to K^+\pi^0\pi^0(\eta)$ with fractions of $r_{K^{*+}\pi^0}$ and $r_{K^{*+}\eta}$ determined above.
The obtained DT efficiencies ($\epsilon_{\rm DT}^i=\epsilon^{i}_{{\rm tag},{\rm sig}}$)
and signal efficiencies ($\epsilon^{i}_{\rm sig}$) for individual decays are summarized in Table~\ref{tab:doubletageff2pipm}.

\begin{table}[htp]
\centering
\small
\caption{ The ST yields ($N_{\rm ST}^{i}$), the ST efficiencies ($\epsilon^{i}_{\rm tag}$),
the DT efficiencies ($\epsilon_{\rm DT}^i=\epsilon^{i}_{{\rm tag},{\rm sig}}$),
and the signal efficiencies ($\epsilon^{i}_{\rm sig}$).
Compared to the mixed signal MC events,
the lower signal efficiencies for the resonant decays are mainly due to that
the $\pi^0$s from $K^{*+}$ decays have much lower momenta and an additional $K^{*+}$ mass requirement.
For $D^-\to K^+\pi^{-}\pi^-\pi^0$, the efficiencies are lower than those of the other two tag modes,
mainly because of more migrations of low momentum pions between tag and signal sides.
The efficiencies do not include the BFs of subresonance decays.
The uncertainties are statistical only.}
\label{tab:doubletageff2pipm}
\begin{tabular}{lccccc}
  \hline\hline
Tag mode $i$            & $D^-\to K^+\pi^-\pi^-$ & $D^-\to K^{0}_{S}\pi^{-}$&$D^-\to K^+\pi^{-}\pi^-\pi^0$ & Average\\ \hline
$N^{i}_{\rm ST}$        & $798935\pm1011$        & $93308\pm329$             &$258044\pm1036$ &... \\
$\epsilon^{i}_{\rm tag}$& $0.5190\pm0.0008$      & $0.5180\pm0.0017$         & $0.2692\pm0.0009$ &... \\ \hline
$\epsilon^{i}_{{\rm tag},{D^+\to K^+\pi^0\pi^0}}$& $0.0966\pm0.0001$      & $0.1004\pm0.0003$   & $0.0429\pm0.0001$     & ... \\
$\epsilon^{i}_{D^+\to K^+\pi^0\pi^0}$            & $0.1862\pm0.0003$      & $0.1937\pm0.0008$   &  $0.1595\pm0.0006$    & $0.1808\pm0.0003$ \\ \hline
$\epsilon^{i}_{{\rm tag},{D^+\to K^{*+}\pi^0}}$&$0.0697\pm0.0001$&$0.0731\pm0.0003$ &$0.0305\pm0.0001$ &... \\
$\epsilon^{i}_{D^+\to K^{*+}\pi^0}$            &$0.1344\pm0.0003$&$0.1411\pm0.0008$ &$0.1133\pm0.0005$&$0.1302\pm0.0003$ \\ \hline
$\epsilon^{i}_{{\rm tag},{D^+\to K^+\pi^0\eta}}$ & $0.1093\pm0.0001$&$0.1122\pm0.0003$ &$0.0494\pm0.0001$&...\\
$\epsilon^{i}_{D^+\to K^+\pi^0\eta}$             & $0.2105\pm0.0004$&$0.2166\pm0.0009$ &$0.1835\pm0.0007$&$0.2050\pm0.0003$ \\ \hline
$\epsilon^{i}_{{\rm tag},{D^+\to K^{*+}\eta}}$& $0.0888\pm0.0002$ & $0.0915\pm0.0006$ &$0.0395\pm0.0001$ &...\\
$\epsilon^{i}_{D^+\to K^{*+}\eta}$            & $0.1710\pm0.0005$ & $0.1769\pm0.0012$ &$0.1467\pm0.0006$ &$0.1660\pm0.0004$ \\ \hline\hline
\end{tabular}
\end{table}

For each signal decay, the statistical significance is evaluated using $\sqrt{-2{\rm ln}(\mathcal L_0/\mathcal L_{\rm max})} $,
where $\mathcal L_{\rm max}$ is the maximum likelihood of the nominal fit
and $\mathcal L_0$ is obtained by refitting the $M_{\rm BC}^{\rm tag}$ versus $M_{\rm BC}^{\rm sig}$ distribution without the signal PDF. Especially, the peaking background of the non-resonant component has been fixed for $D^+\to K^{*+}\pi^0$ and $D^+\to K^{*+}\eta$.
The resulting statistical significances are $8.8\sigma$, $5.5\sigma$, $2.7\sigma$, and $3.2\sigma$ for $D^+\to K^+\pi^0\pi^0$, $D^+\to K^+\pi^0\eta$, $D^+\to K^{*+}\pi^0$, and $D^+\to K^{*+}\eta$, respectively.
In addition, 10000 toy MC studies show that the 2D fit is stable and no potential bias is found for each signal decay.

The measured values for ${ N_{{\rm DT}}}$, $\epsilon^{}_{{\rm sig}}$, and $\mathcal B_{\rm sig}$ are summarized in Table~\ref{tab:DT}.
Because there is no significant signal for $D^+\to K^{*+}\pi^0$, we set an upper limit on its decay BF at the 90\% confidence level to be $5.4\times10^{-4}$. This is set utilizing the Bayesian approach after incorporating the associated systematic uncertainty~\cite{li}, as discussed later.

\begin{table}[htp]
\centering
\caption{\label{tab:DT}
\small
The DT yields in data~($N_{\rm DT}$), the signal efficiencies ($\epsilon_{\rm sig}$ = $\epsilon_{\rm DT}^i/\epsilon_{\rm ST}^i$ ) and the obtained BFs.
The first and second uncertainties are statistical and systematic, respectively. The efficiencies do not include the BFs of $\pi^0$, $\eta$ and $K^{*+}$ decays.
The lower efficiency for $D^+\to K^+\pi^0\pi^0$ is mainly due to the $K^0_S$ rejection.}
\begin{tabular}{lcccc}
\hline\hline
Decay mode  & $N_{\rm DT}$ & $\epsilon_{\rm sig}$\,(\%)& $\mathcal B_{\rm sig}~(\times\,10^{-4})$\\ \hline
$D^+\to K^+\pi^0\pi^0$   &$42.8\pm7.2$&$18.08\pm0.03$&$2.1\pm0.4\pm0.1$\\
$D^+\to K^+\pi^0\eta$    &$19.2\pm5.0$&$20.50\pm0.03$&$2.1\pm0.5\pm0.1$\\
$D^+\to K^{*+}\pi^0$    &$16.6^{+6.6}_{-6.2}$&$13.02\pm0.03$&$3.4^{+1.4}_{-1.3}\pm0.1$ \\
$D^+\to K^{*+}\eta$     &$10.9^{+4.4}_{-3.8}$&$16.60\pm0.04$&$4.4^{+1.8}_{-1.5}\pm0.2$\\
\hline\hline
\end{tabular}
\end{table}

\section{SYSTEMATIC UNCERTAINTY}

One of the advantages of the DT method is that most of the uncertainties associated with the ST selection cancel.
The systematic uncertainties in the BF measurements are mainly from the following sources.
They are reported relative to the measured BFs.
\begin{itemize}
\item {\bf ST yields ($N_{\rm tag}$):} The uncertainty of the total ST $D^-$ yield, which
is mainly due to the fit to the $M_{\rm BC}^{\rm tag}$ distribution,
has been previously estimated to be 0.5\% in Ref.~\cite{bes3-etaetapi}.
\item {\bf $K^\pm$ tracking or particle identification (PID):} The efficiencies of tracking and PID of the $K^+$ are studied with DT $D\bar D$ hadronic events.
The systematic uncertainty for $K^+$ tracking and PID is 1.0\% for each.
\item {\bf $\pi^0\,(\eta)$ reconstruction:} The efficiency of $\pi^0$ reconstruction is investigated using DT $D\bar D$ hadronic decay samples of $D^0\to K^-\pi^+$, $K^-\pi^+\pi^+\pi^-$ versus $\bar D^0\to K^+\pi^-\pi^0$, $K^0_S\pi^0$~\cite{epjc76,cpc40}. The systematic uncertainty due to $\pi^0$ reconstruction is $2.0\%$ per $\pi^0$.
Based on the $\pi^0$ uncertainty, the systematic uncertainty of $\eta$ reconstruction is also taken to be 2.0\%.
The total systematic uncertainty due to $\pi^0\pi^0$ or $\pi^0\eta$ reconstruction is obtained to be 4.0\% by adding each of them linearly.
\item {\bf Quoted BFs:} The uncertainties on the quoted BFs of $\eta\to\gamma\gamma$ and $\pi^0\to\gamma\gamma$ are 0.5\% and 0.03\%~\cite{pdg2020}, respectively.
\item {\bf 2D fit:} The systematic uncertainty of the 2D fit is mainly due to the signal and background shapes.
 To compensate for the possible data-MC difference of the signal, the MC-simulated signal shapes have been smeared by a Gaussian resolution function with parameters derived from the control sample of $D^+\to \pi^+\pi^0\pi^0$.
 Therefore, the systematic uncertainty due to the signal shape is ignored.
To consider the uncertainty of center-of-mass energy calibration~\cite{calibration},
the endpoint of the ARGUS background function is varied by $\pm0.2$ MeV/$c^2$.

The changes of the BFs are assigned as the corresponding systematic uncertainties,
which are 0.2\% for both $D^+\to K^+\pi^0\pi^0$ and $D^+\to K^+\pi^0\eta$, but are negligible for $D^+\to K^{*+}\pi^0$ and $D^+\to K^{*+}\eta$.

\item {\bf $D^+D^-$ opening angle:} The systematic uncertainty arising from the $D^+D^-$ opening angle requirement
is studied by using the control sample of $D^+\to \pi^+\pi^0\pi^0$.
The difference of the acceptance efficiencies between data and MC simulation, 1.2\%, is assigned as the corresponding systematic uncertainty.
\item {\bf $\Delta E^{\rm sig}$ requirement:} The systematic uncertainty of the $\Delta E_{\rm sig}$ requirement is estimated by
convolving with one Gaussian resolution function obtained from the control sample with the $\Delta E_{\rm sig}$ distribution of the signal MC events.
The change of the DT efficiency is found to be negligible. Therefore, the corresponding systematic uncertainty is neglected.
\item {\bf $K^0_S$ rejection:} The systematic uncertainty due to the $K^0_S$ rejection is also negligible
since the BFs are found to be insensitive to
shrinking or enlarging the $K^0_S$ rejection window by $0.02$ GeV/$c^2$, which is about two standard deviations of the fitted $K^0_S(\to \pi^0\pi^0)$ resolution, and
taking into account correlations of the two signal samples with the nominal and varied $K^0_S$ signal regions~\cite{uncer}.
\item {\bf $K^{*+}$ signal region:} The systematic uncertainty of the $K^{*+}$ signal region is studied using DT events from the processes $D^0\to K^-\pi^+$ and $K^-\pi^+\pi^0$ versus $\bar D^0\to K^{*+}(\to K^+\pi^0)e^- \bar \nu_e$.
The change of the DT efficiencies after convolving with the obtained Gaussian resolution function with the $M_{K^+\pi^0}$ distributions, 0.1\%, is assigned as the associated uncertainty.
\item {\bf MC statistics:} The uncertainties of MC statistics are 0.2\%, 0.2\%, 0.3\%, and 0.3\%
for $D^+\to K^+\pi^0\pi^0$, $D^+\to K^+\pi^0\eta$, $D^+\to K^{*+}\pi^0$, and $D^+\to K^{*+}\eta$, respectively.
\item {\bf MC modeling:} The systematic uncertainties related to the MC modeling
for $D^+\to K^+\pi^0\pi^0$ and $D^+\to K^+\pi^0\eta$ are estimated by varying $r_{K^{*+}\pi^0(\eta)}$ by $\pm 1\sigma$.
The changes of the detection efficiencies are assigned as the corresponding systematic uncertainties,
which are 2.1\% and 1.6\% for $D^+\to K^+\pi^0\pi^0$ and $D^+\to K^+\pi^0\eta$, respectively.
\item {\bf Scale factor of $K^{*+}$ sideband:}
The systematic uncertainties due to the scale factors of $K^{*+}$ sideband
are examined by varying $f_{K^{*+}\pi^0}$ and $f_{K^{*+}\eta}$ by $\pm 1\sigma$.
The changes of the re-measured BFs, 1.4\% and 0.5\%, are assigned as
the systematic uncertainties for $D^+\to K^{*+}\pi^0$ and $D^+\to K^{*+}\eta$, respectively.
\item {\bf Multiplicities of tag and signal sides:}
To verify the smallest $|\Delta E|$ selection method, we have examined the multiple candidate rates for the tag and signal sides.
Due to limited signal statistics, the signal side is examined with the control sample of $D^+\to \pi^+\pi^0\pi^0$,
which has similar multiple candidate rates as our signal candidates.
The multiple candidate rates of $D^-\to K^+\pi^-\pi^-$, $D^-\to K^0_S\pi^-$, $D^-\to K^+\pi^-\pi^-\pi^0$, and $D^+\to \pi^+\pi^0\pi^0$
are about 0.4\%, 0.2\%, 9.9\%, and 1.7\% with negligible uncertainties, respectively, for both data and MC simulation. Therefore, the relevant effect is ignored in this analysis.
\end{itemize}

Adding the above effects in quadrature yields the total systematic uncertainty for each signal process.
They are 4.7\%, 4.5\%, 4.4\%, and 4.3\% for $D^+\to K^+\pi^0\pi^0$, $D^+\to K^+\pi^0\eta$, $D^+\to K^{*+}\pi^0$, and $D^+\to K^{*+}\eta$, respectively.
Table~\ref{tab:relsysuncertainties} summarizes the systematic uncertainties discussed above.

\begin{table}[htp]
\centering
\small
\caption{
Systematic uncertainties (\%) in the measurements of the BFs.}
\label{tab:relsysuncertainties}
\centering
\begin{tabular}{ccccc}
  \hline\hline
Uncertainty&$K^+\pi^0\pi^0$&$K^+\pi^0\eta$&$K^{*+}\pi^0$&$K^{*+}\eta$\\
  \hline
$N_{\rm tag}$                      &0.5&0.5&0.5&0.5\\
$K^\pm$ tracking                   &1.0&1.0&1.0&1.0\\
$K^\pm$ PID                        &1.0&1.0&1.0&1.0\\
$\pi^0\,(\eta)$ reconstruction     &4.0&4.0&4.0&4.0\\
Quoted BFs                         &Negligible&0.5&Negligible&0.5\\
2D fit                             &0.2&0.2&Negligible&Negligible\\
$D^+D^-$ opening angle             &1.2&1.2&1.2&1.2\\
$\Delta E^{\rm sig}$ requirement           &Negligible&Negligible&Negligible&Negligible\\
$K^0_S$ rejection                  &Negligible&Negligible&Negligible&Negligible\\
$K^{*+}$ signal region             &-&-&0.1&0.1\\
MC statistics                      &0.2&0.2&0.3&0.3\\
MC modeling                        &2.1&1.6&-&-\\
Scale factor of $K^{*+}$ sideband   &-&-&1.4&0.5\\
Multiplicities of tag and signal sides&Negligible&Negligible&Negligible&Negligible\\
\hline
Total                              &4.7&4.5&4.4&4.3\\
\hline\hline
\end{tabular}
\end{table}

\section{SUMMARY}

In summary, using 2.93\,fb$^{-1}$ of $e^+e^-$ annihilation data~\cite{lum_bes31,lum_bes32} taken at $\sqrt s=$ 3.773~GeV, we report the first observations of the DCS decays $D^+\to K^+\pi^0\pi^0$ and $D^+\to K^+\pi^0\eta$, as well as the first searches for $D^+\to K^{*+}\pi^0$ and $D^+\to K^{*+}\eta$. It should be noted that the BFs of $D^+\to K^{*+}\pi^0(\eta)$ are measured under the assumptions that there is no interference between resonant and non-resonant components, and the non-resonant component is uniformly distributed in the phase space. The obtained BFs are summarized in Table \ref{tab:DT}.
We also set an upper limit on the BF of $D^+\to K^{*+}\pi^0$ decays of  to be $5.4\times10^{-4}$ at the 90\% confidence level.
Our $D^+\to K^{*+}\pi^0$ and $D^+\to K^{*+}\eta$ BF results supply important information for more detailed investigations of SU(3)-flavor symmetry breaking effects as well as for the understanding of $CP$ violation phenomena in hadronic decays of charmed mesons.
Our measured ${\mathcal B}_{D^+\to K^{*+}\pi^0}$ is consistent with the predictions in Refs.~\cite{yufs,theory_2,theory_3},
while ${\mathcal B}_{D^+\to K^{*+}\eta}$ differs from all predictions by approximately $2\sigma$.
With the obtained BFs of $D^+\to K^{*+}\pi^0$ and $D^+\to K^{*+}\eta$, we set an upper limit on the BF ratio to be ${\mathcal B}_{D^+\to K^{*+}\pi^0}/{\mathcal B}_{D^+\to K^{*+}\eta}<1.64$ at the 90\% confidence level.
Combining our ${\mathcal B}_{D^+\to K^+ \pi^0\pi^0}$ and ${\mathcal B}_{D^+\to K^+ \pi^0\eta}$ with ${\mathcal B}_{D^+\to K^- \pi^+\pi^+}=(9.38\pm0.16)\%$~\cite{pdg2020} and ${\mathcal B}(D^+\to K^0_S \pi^+\eta)=(1.31\pm0.04\pm0.03)\%$~\cite{papnew1}, we obtain the relative DCS to CF BF ratios
${\mathcal B}_{D^+\to K^+\pi^0\pi^0}/{\mathcal B}_{D^+\to K^-\pi^+\pi^+}=(2.24\pm0.40)\times 10^{-3}$ and
${\mathcal B}_{D^+\to K^+\pi^0\eta}/{\mathcal B}_{D^+\to \bar K^0\pi^+\eta}=(8.01\pm1.97)\times 10^{-3}$.
They correspond to $(0.77\pm0.14)\tan^4\theta_C$ and  $(2.64\pm0.68)\tan^4\theta_C$, respectively.
The former ratio is consistent with the naive prediction $\frac{2}{3}\tan^4\theta_C$, while the latter differs from
the naive expectation of $\tan^4\theta_C$ by $2.4\sigma$.
Making use of ${\mathcal B}_{D^+\to K^+ \pi^+\pi^-}=(4.91\pm0.09)\times 10^{-4}$ \cite{pdg2020},
we determine ${\mathcal B}(D^+\to K^+ \pi^0\pi^0)/{\mathcal B}(D^+\to K^+ \pi^+\pi^-) = 0.43 \pm 0.08$,
which is consistent with prediction assuming isospin symmetry between these two decays.

\section{ACKNOWLEDGEMENT}
The authors thank Professors Yu-Kuo Hsiao, Qin Qin, and Fusheng Yu for helpful discussions.
The BESIII collaboration thanks the staff of BEPCII and the IHEP computing center for their strong support. This work is supported in part by National Key R\&D Program of China under Contracts Nos. 2020YFA0406400 and 2020YFA0406300; National Natural Science Foundation of China (NSFC) under Contracts Nos. 11775230, 11625523, 11635010, 11735014, 11822506, 11835012, 11935015, 11935016, 11935018, 11961141012, 12022510, 12025502, 12035009, 12035013, 12061131003, 12192260, 12192261, 12192262, 12192263, 12192264, 12192265; the Chinese Academy of Sciences (CAS) Large-Scale Scientific Facility Program; Joint Large-Scale Scientific Facility Funds of the NSFC and CAS under Contracts Nos. U1732263, U1832207, U1932102; CAS Key Research Program of Frontier Sciences under Contract No. QYZDJ-SSW-SLH040; the CAS Center for Excellence in Particle Physics (CCEPP); 100 Talents Program of CAS; INPAC and Shanghai Key Laboratory for Particle Physics and Cosmology; ERC under Contract No. 758462; European Union Horizon 2020 research and innovation programme under Contract No. Marie Sklodowska-Curie grant agreement No 894790; German Research Foundation DFG under Contracts Nos. 443159800, Collaborative Research Center CRC 1044, FOR 2359, FOR 2359, GRK 214; Istituto Nazionale di Fisica Nucleare, Italy; Ministry of Development of Turkey under Contract No. DPT2006K-120470; National Science and Technology fund; Olle Engkvist Foundation under Contract No. 200-0605; STFC (United Kingdom); The Knut and Alice Wallenberg Foundation (Sweden) under Contract No. 2016.0157; The Royal Society, UK under Contracts Nos. DH140054, DH160214; The Swedish Research Council; U. S. Department of Energy under Contracts Nos. DE-FG02-05ER41374, DE-SC-0012069.

\end{document}